# Use of Electrical Resistivity Tomography to Map the Tree Roots

Xiaolong Liang[1], Xiao-ping Wu[1]

[1]School of Earth and Space Sciences, University of Science and Technology of China, Hefei 230026, China

**Abstract**: An efficient advanced numerical model for mapping the distribution of the buried tree roots is presented. It not only simplify the complicate root branches to an easy manipulated model, but also grasp the main structure of tree roots ignoring the unnecessary minutiae, and thus provide an intuitive impression of subsurface invisible anomalies. The processing model is combined with an adaptive finite element method, which can automatically generate unstructured triangular meshes during the process of discretization, which also enable user to specifically set the resistivity along each part of tree roots.

## Introduction

Electrical resistivity tomography (ERT), which provides a two-dimensional depth section or modeled pseudosection across a survey area, accurately records depth information and assesses deeply buried anomalous bodies in rough outline. ERT survey based on geophysical measurements of resistivity variations in subsurface, the principle of which by applying the electrical electrodes across a constant current through the underground medium[1], the measured data are processed to obtain the relevant position and resistivity of subsurface electrical layers.

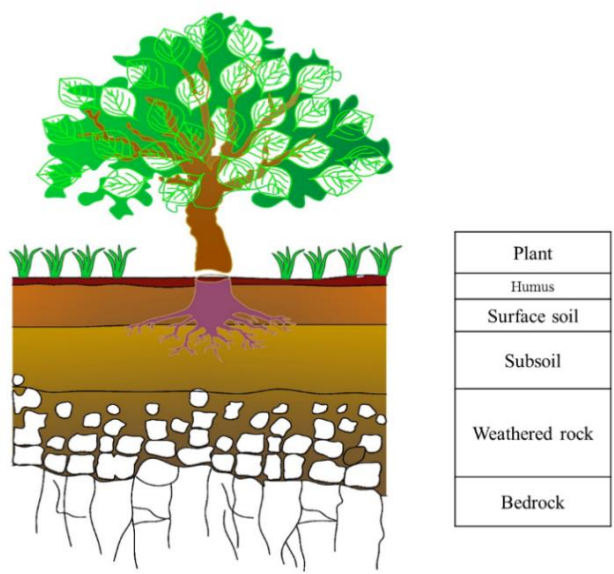

**Fig.1. Sketch of the distribution of tree roots buried in subsurface**

ERT also can be used in geophysical investigation associated with landscapes which are not generally amenable for other types of survey such as the tropical rain forests, the temperate grasslands and wetlands, agricultural lands and areas of lush vegetation. Though traditional techniques used for detect root zones (soil cores, texture and layers) provide relatively accurate information, while their destructive, time consuming and labor-intensive limited the application in a large scale of field work respect to the soil volume and surface area. For more deeply understanding of plant growth, root biomass, soil texture, and all of those important functions for climatic monitoring and environmental protection [2]. As a convenient, nondestructive technique, it is imperative to exactly map the tree roots zone using the numerical simulation and experimental techniques of electrical resistivity tomography in 2D.

Fig.1 shows a sketch of tree roots buried in different soil layers: from top to bottom, followed by low understory plants, leached layer with humus and leaf litter, surface soil, subsoil, weathered rock (produce soil parent materials) and bedrock (parent rock). It is noticeable that the tree roots mainly distributed in surface soil and subsoil. Indeed, the sum of all the roots of a plant, called root system. Roots can be divided into straight root system (dicotyledonous plants) and fibrous root system (monocotyledonous plants), and the most tree roots belonging to straight root system.

Under normal circumstances, the horizontal distribution of tree roots generally consistent with the majority size of tree's crown. Under suitable soil conditions, most of tree roots in the vertical direction are concentrated in the range of 40-80 cm deep in the underground, while roots with absorbing function are located about 20 cm deep in the soil, but different tree roots in depth distribution vary greatly.

## Method

Soil in the field is typically a heterogeneous and lossy complex media [3], and its properties vary greatly and change frequently with external environmental parameters. The electrical properties of tree root zone mainly derive from dissolved ions in fluid or sap content, wooden structure, soft composition, and anisotropic cells capability to conduct currents. Tree root zones mainly being filled by mixed root networks and surrounding invisible materials. Thus, the important task urgently is simplified the sophisticated actual distribution to an intuitive succinct model (Fig.2), which achieved in using a Matlab package (IP4DI) made by Karaoulis[4].

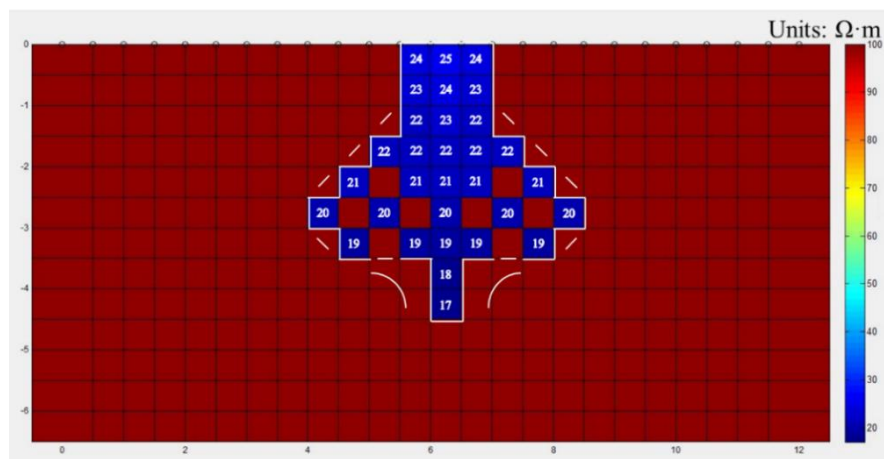

**Fig.2. Cross-section of simplified tree roots model with relevant resistivity in different branches**
(20±5Ω•m, from top tree stump<25Ω•m> to bottom root tip<17Ω•m>)

The model data acquisition based on dipole-dipole array, utilizing 25 electrodes with 0.25m electrode space, but there exist one inconvenient function that the resistivity of model's each grid must be set manually one by one, or it may be developed for the purpose of much more self-control and self-design.

Fig.2 shows the main structure of the simulated tree roots model, the number in blue grids represent the resistivity of different part of root branches, the law of resistivity variation (20±5Ω•m) along the main root and lateral roots based on the actual biological characteristics (Young and soft fine roots absorb water and dissolved fertilizer, lateral roots resistivity relatively low. Older and wooden coarse roots partly transport sap and nutrient, main root resistivity relatively high). The model assumed this tree roots buried in a homogeneous isotropic sandy clay (red grids) with the resistivity of 100Ω•m.

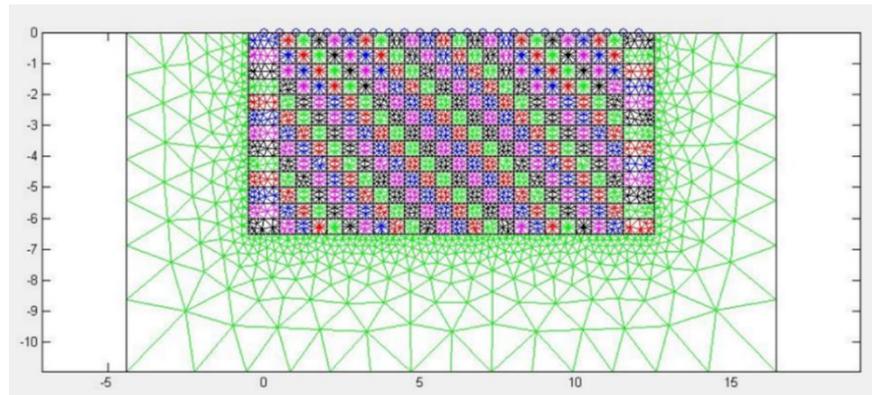

**Fig.3. Automatic generation of unstructured triangular meshes for general 2D geometry**

After forwarding the tree roots model, simulated electric field data being generated and re-read to be processing for discretization. In Fig.3, a discretized subsurface model scenario for the 2D roots zone is obtained from numerical simulation using an adaptive refined finite-element method [5].

Then modified relevant parameters of inversion, such as iteration numbers, inversion type, jacobian type, boundary condition and so on. For the simplified tree roots model, using the Occam's inversion algorithm and full jacobian (calculate jacobian in every iteration) to calculate the inversion results, finally mapping the image. Through several statistical analysis, in general, the inversion process saving time and convergent rapidly (5~7 iterations).

**Results**

The model separately simulated both cases of simplified tree roots and added noise root zones in a homogeneous isotropic sandy clay. These two cases are assumed that resistivity of 100Ω•m for the sandy clay, 20±5Ω•m for the resistive root zones, respectively.

Fig.4 provide a sequence of snapshots, every iteration in the process of inversion (above the green line), and a comparison between the simplified tree roots model and the convergence image (above the green line). Iteration number and RMS error at the top of image, left and bottom of image paste the depth and distance(m), right is the color bar (log10ρ, Ω•m) show approximate range of resistivity from 20 to 125Ω•m.

With the increasing of iteration number, the EMS error decreases rapidly, and the inversion region gradually accelerate the speed of convergence whose outline approaching root zone. Until automatically termination of the inversion, this method just expend 6 iterations within 10 minutes.

It can be found from comparison that final inversion result similar with simplified tree roots model, especially in outline (contained in white/black dashed box) and relative position (constrained by white/black vertical and horizontal marking arrow) of roots zone. Although the bottom and the minor details of root branches are not clearly enough, but it still exhibit a rough range of the buried tree roots. Therefore the model can be used as a reference for precious and rare trees transplant, and also satisfy the interest for buildings and avenue which suffer subsidence or structural damage from nearby trees.

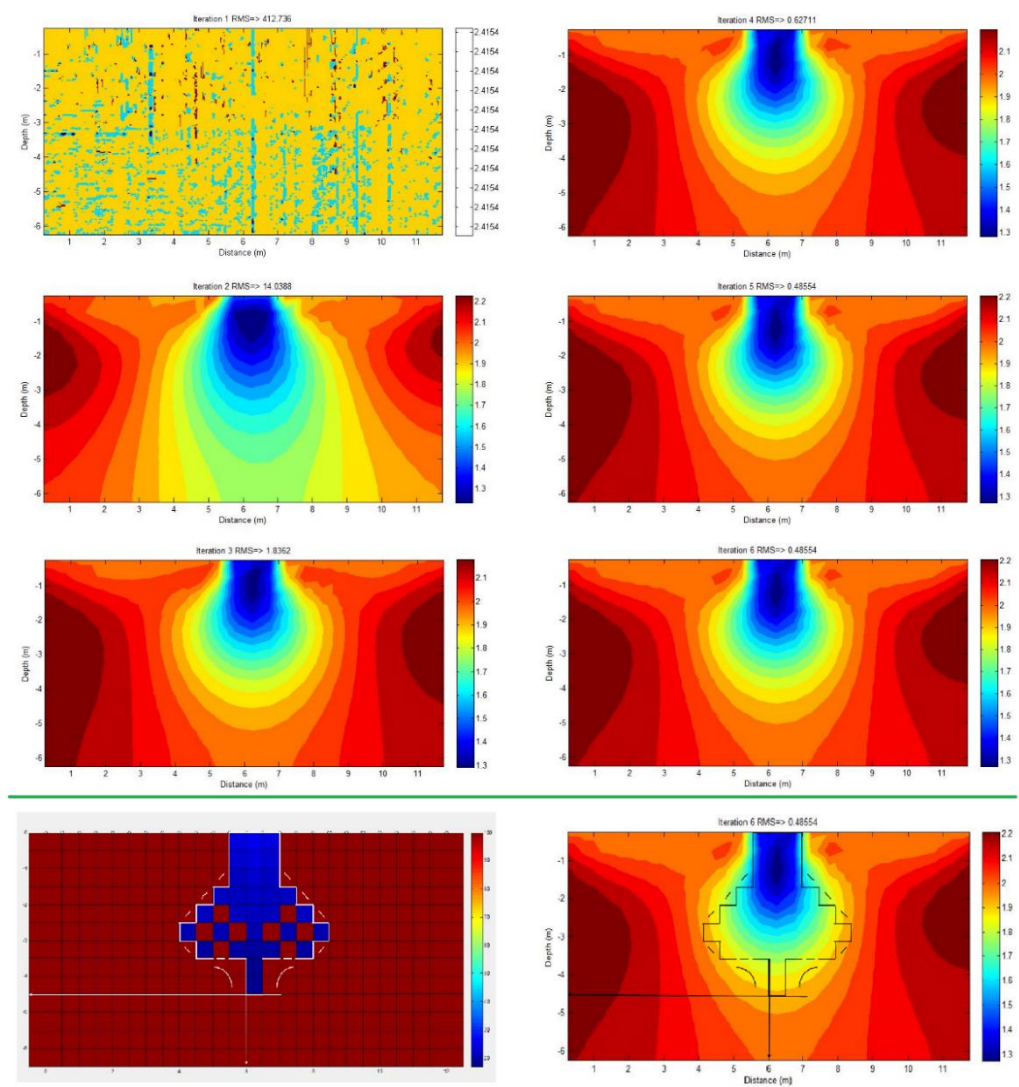

**Fig.4. Inversion process snapshots of simplified tree roots model with Occam's inversion algorithm (above the green line). Comparison between tree roots model and the inversion results (beneath the green line).**

In order to obtain more realistic model, it is a common used method that added noise to the original forwarding data before inversion starting. Noised data achieved from dot products which composed of random number between 0 and 1, original data and noise ratio. Noise added equation calculated as follows:

**Noised Data = Original Data + ( Original Data .* Rand (0, 1) ) .* Noise Ratio**

Noise ratio need to be multiplied to the original data: 0.01, 0.02, 0.04, 0.06, 0.08, 0.10, 0.15, 0.20

In this test, keeping the majority of parameters and constrained condition same as the simplified tree roots model mentioned above, apart from changing algorithm from Occam's inversion to Gauss-Newton inversion. As can be seen from Fig.5, despite all of 8 images added different magnitude noise to same original model, they almost show any noticeable change among the tree root zone. This comparison indicates that the model has a too unresponsive reaction to the external environment impact, and suggests that this model still needs to be modified and corrected for the possibility of practical application.

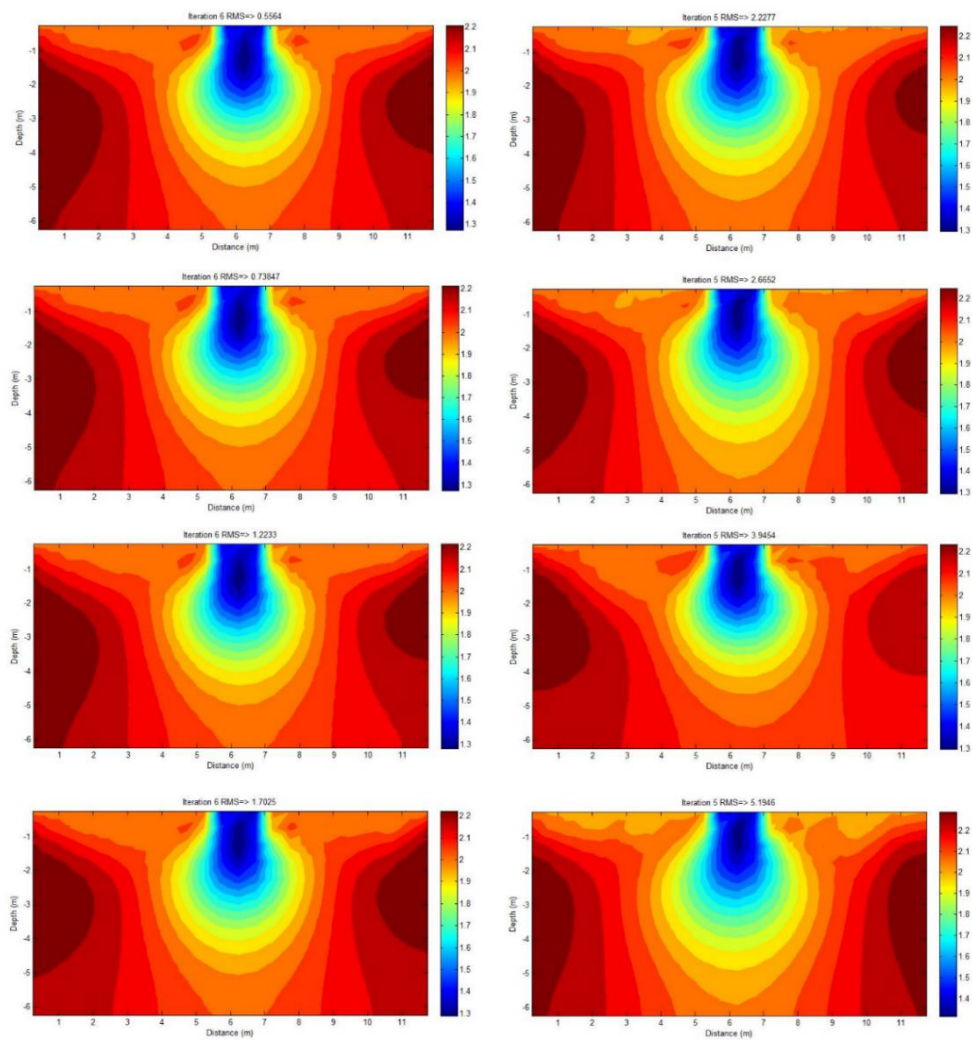

**Fig.5. Comparison of inversion results with same original model added different noise**
(Snapshots arranged as an inverted 'N' type. Noise ratio: 0.01, 0.02, 0.04, 0.06, 0.08, 0.10, 0.15, 0.20)

## Conclusion

In the actual field, beneath root branches are really very complex. Unlike the model controlled laboratory condition, anisotropic heterogeneous soils, adjacent plant's living or dead roots, caves made by insects and little creatures as well as many other uncontrollable factors, all of them have an adverse effect on apparent resistivity data acquisition. However, for present, ERT still can be used as an easy, save time and efficient approach, especially in describing shape and behavior of tree roots in the subsurface.

## Acknowledgments

The paper is sponsored by National Natural Science Foundation of China (41074048, 41130420), National High Technology Research and Development Program of China (863 Program) (2012AA061403, 2012AA09A201), and Special-funded programme on national key scientific instruments and equipment development (2011YQ05006008).